\documentclass[conference,twocolumn]{IEEEtran}
\usepackage{}
\usepackage{stfloats}
\usepackage{cite}
\usepackage[cmex10]{amsmath}
\usepackage{balance}
\usepackage{tabularx,colortbl}
\definecolor{Gray}{gray}{0.9}

\usepackage{amsmath,amssymb,amsthm,mathrsfs,dsfont}
\usepackage{graphicx} 

\usepackage{amsmath}
\usepackage{mathtools}
\usepackage{amssymb}

\usepackage{booktabs}
\usepackage{threeparttable}
\usepackage{multirow}
\usepackage{epstopdf}
\usepackage{algorithm}
\usepackage{algpseudocode}

\usepackage{subfigure}
\usepackage{stfloats}
\usepackage{bm}

\usepackage{verbatim}
\usepackage{float}
\usepackage{textcomp}
\usepackage{mathrsfs}
\usepackage{amsmath}
\usepackage{amsthm}
\usepackage{amssymb}
\usepackage{graphicx}
\usepackage{setspace}

\IEEEoverridecommandlockouts

\begin{document}

\title{What is the Optimal Network Deployment \\
for a Fixed Density of Antennas?}

\author{{\normalsize{}Xuefeng Yao$^{\P}$, Ming Ding$^{\ddagger}$, David L$\acute{\textrm{o}}$pez
P$\acute{\textrm{e}}$rez$^{\dagger}$}\emph{\normalsize{}, }{\normalsize{}Zihuai Lin$^{\P}$, Guoqiang
Mao$^{\nparallel}{}^{\dagger}$}\emph{\normalsize{}}\\
\textit{\small{}$^{\P}$School of Electrical and Information Engineering,
University of Sydney, Australia }{\small{}\{xuefeng.yao@sydney.edu.au\}}\textit{\small{}}\\
\textit{\small{}$^{\ddagger}$Data61, CSIRO, Australia }{\small{}\{Ming.Ding@data61.csiro.au\}}\textit{\small{}}\\
\textit{\small{}$^{\dagger}$Nokia Bell Labs, Ireland }{\small{}\{david.lopez-perez@nokia.com\}}\textit{\small{}}\\
\textit{\small{}$^{\nparallel}$School of Computing and Communication,
University of Technology Sydney, Australia}
}

\maketitle
\begin{abstract}
In this paper, we answer a fundamental question:
when the total number of antennas per square kilometer is fixed,
what is the optimal network deployment?
A denser network with a less number of antennas per base station (BS) or the opposite case.
To evaluate network performance,
we consider a practical network scenario with a fixed antennas density and multi-user multiple-input-multiple-output (MU-MIMO) operations for single-antenna users.
The number of antennas in each BS is calculated by dividing the antenna density by the BS density.
With the consideration of several practical network models,
i.e., pilot contamination,
a limited user equipment (UE) density and probabilistic line-of-sight (LoS)/non-line-of-sight (NLoS) path loss model,
we evaluate the area spectral efficiency (ASE) performance.
From our simulation results,
we conclude that there exists an optimal BS density for a certain UE density to maximize the ASE performance when the antenna density is fixed.
The intuition is that
\emph{(i)} by densifying the network with more BSs,
we can achieve a receive power gain due to the smaller distance between the typical UE and its serving BS;
\emph{(ii)} by installing more antennas in each BS,
we can achieve a beamforming gain for UEs using MU-MIMO,
although such beamforming gain is degraded by pilot contamination;
\emph{(iii)} thus, a trade-off exists between the receive power gain and the beamforming gain,
if we fix the antenna density in the network.
\end{abstract}

%

\IEEEpeerreviewmaketitle

\section{Introduction}

Mobile data traffic is predicted to grow 1000x from now until 2030~\cite{cisco},
and dense small cell networks (SCNs) and massive multiple input multiple output (mMIMO) are considered the major pilar technologies to meet this ever-increasing capacity demand in the years to come~\cite{Tutor_smallcell,Massive}.

From 1950 to 2000,
network capacity was dramatically increased through network densification by a factor of 2700x~\cite{cisco}.
During the first decade of 2000,
the 3rd Generation Partnership Project (3GPP) standardised the 4th-Generation (4G) Long Term Evolution (LTE) systems,
which kept paying special attention to network densification and small cells,
as an effective approach to increase capacity~\cite{Tutor_smallcell}.
The first standardisation efforts on New Radio (NR) also indicate that network densification will remain as one of the workhorses in 5G systems.
Small cells are a powerful approach to fuel fast-growing network demand due to its fundamental benefits.
However, how to deploy them in a cost-effective manner has been a big concern for both vendors and operators,
and thus of the research community.
A good understanding of the performance of dense SCNs is necessary.

Up to now,
many studies on dense SCNs have focused on deriving the area spectral efficiency (ASE) performance.
However, most of them only consider a limited number of factors.
In~\cite{jeffery},
the authors considered a single-slope path loss model without differentiating line-of-sight (LoS) and non-LoS (NLoS) transmissions.
In~\cite{losNlos},
the authors embraced LoS and NLoS transmissions,
but only considered an infinity UE density and a single-input single-output (SISO) system.
In~\cite{PLmodel},
the authors further included in the analysis a finite UE density and an idle mode capability (IMC),
which is key to achieve a good performance of dense SCNs.
In~\cite{mimohet},
multiple antennas were considered while analysing coverage probability.

In addition to dense SCNs,
mMIMO,
considered as a scaled-up version of multi-user MIMO (MU-MIMO),
also has the potential to further increase network capacity by exploiting the degrees of freedom in the spatial domain.
Indeed, mMIMO has already been adopted as a main technology to improve ASE in 5G systems~\cite{Massive}.
It is important to note that the larger the number of antennas, the larger the number of degrees of freedom, and thus the more multiplexing opportunities.
However, when time division duplex (TDD) systems are considered,
due to a finite channel coherent time,
the performance of mMIMO systems may be limited by inaccurate channel state information (CSI).
Pilot contamination is considered as a major bottleneck,
which occurs when the same set of uplink training sequences is reused across neighbouring cells~\cite{MIMO2}.
Other channel estimation impairments also play role.

Looking at mMIMO deployment aspects,
in~\cite{MIMO2},
the authors showed that a better performance can be achieved by increasing the number of antennas at the BS and using a simple signal processing.
In~\cite{Massive2016},
the authors analysed the uplink signal to interference plus noise ratio (SINR) and rate performance in a mMIMO system,
considering a single-slope path loss model,
without differentiating LoS and NLoS transmissions.
In~\cite{hetMassivemimo},
the authors derived the downlink achievable rate in mMIMO heterogeneous cellular networks,
while accounting for LoS and NLoS transmissions and an infinity UE density.
It is important to note that pilot contamination was considered in the above three studies.
However, little work has been done on understanding the impact of the BS and UE densities in the network capacity performance.

Generally speaking and for a certain UE density,
the more BSs and/or the more antennas per BS,
if operated appropriately,
the higher the network capacity~\cite{Ding2017varFactors}.
However, up to now,
it is unclear what is the optimal network deployment,
i.e., the optimal combination of the number of BSs and antennas per BS in a given area,
when the antenna density (antennas/km$^2$) is fixed.
In particular, two extreme cases are of great interest:
\emph{i)} a dense SCN network where all BSs have a single antenna, and
\emph{ii)} a sparse mMIMO network where many antennas are concentrated in a few BSs.
Which network deployment is better in terms of the network capacity performance?

In this paper, we give a first answer to this fundamental question by means of computer simulations.
From our simulation results,
we conclude that, for a certain UE density, there exists an optimal BS density to maximise the ASE performance when the antenna density is fixed.
The intuition is that
\emph{(i)} by densifying the network with more BSs,
we can achieve a receive power gain due to the smaller distance between the typical UE and its serving BS;
\emph{(ii)} by installing more antennas in each BS,
we can achieve a beamforming gain for UEs using MU-MIMO,
although such gain will be degraded by pilot contamination.
Thus, a trade-off exists between the receive power gain and the beamforming gain,
if we fix the antenna density in the network.

The rest of this paper is structured as follows.
Section~II describes the network scenario and the wireless system model considered in this paper.
Section~III presents our numerical results, with remarks shedding new light on the trade-off between the receive power gain and the beamforming gain,
if we fix the antenna density in the network.
Section~IV draws the conclusions.

$\textbf {Notations}$: We use $\bf A$, $\bf a$, and $a$ to denote a matrix, a vector and a scalar, respectively.
${\bf A}^{T}$ and ${\bf A}^{H}$ represent the transpose and conjugate transpose of $\bf A$, respectively.
${\mathbb C}^{m \times n}$ denotes a set of complex numbers with a dimension of $m \times n$.
$\cal A$ denotes a set,
while ${\cal {NB}}(m,n)$ represents a negative binomial distribution with parameters $m$ and $n$.

\section{system model}

In this section, we present the network scenario, wireless system model and pilot-aided MIMO channel estimation considered in this paper.

\subsection{Network Scenario}

We consider a downlink (DL) cellular network with BSs deployed on a plane according to a homogeneous Poisson point process (HPPP) $\Phi$
with a density of $\lambda$ $\textrm{BSs/km}^{2}$.

Active DL UEs are also Poisson distributed in the considered network with a density of $\rho$ $\textrm{UEs/km}^{2}$.
Here, we only consider active UEs in the network because non-active UEs do not trigger any data transmission.
The typical UE $U_{1}$ is deployed at the origin and its serving BS is denoted as $B_{1}$.

Each BS is equipped with $M$ antennas,
and has a total transmit power of $P_{\textrm{b}}^{\textrm{tx}}$,
where $M$ is calculated as the antenna density divided by the BS density $\lambda$.
Each UE is equipped with a single antenna.

In practice, a BS will enter into idle mode,
if there is no UE connected to it,
which reduces the interference to UEs in neighbouring BSs as well as the energy consumption of the network.
Since UEs are randomly and uniformly distributed in the network,
the active BSs also follow another HPPP distribution $\tilde{\Phi}$~\cite{lambdadynOnPL},
the density of which is $\tilde{\lambda}$ $\textrm{BSs/km}^{2}$.
Note that $\tilde{\lambda}\leq\lambda$ and $\tilde{\lambda}\leq\rho$,
since one UE is served by at most one BS.
Also note that a larger $\rho$ results in a larger $\tilde{\lambda}$.

From~\cite{lambdadynOnPL},
$\tilde{\lambda}$ can be calculated as%
\begin{equation}
	\tilde{\lambda}=\lambda\left[1-\frac{1}{\left(1+\frac{\rho}{q\lambda}\right)^{q}}\right],\label{eq:lambda_tilde_Huang}
\end{equation}
where an empirical value of 3.5 was suggested for $q$~\cite{lambdadynOnPL}.

In this paper,
we consider a pilot-aided channel estimation scheme,
and assume imperfect channel state information (CSI) caused by pilot contamination.
In an uplink training stage,
a scheduled UE transmits a randomly assigned pilot sequence $t_k$ from the set of available training sequences $\cal{T}$.
UEs in each BS reuse the same set of pilot sequence,
i.e. the pilot reuse factor can be as high as one.
However, it should be noted that the pilot reuse factor at a particular time instant strongly depends on the number of pilot sequences and the number of UEs per BS.
For example, a higher number of UEs per BS implies a larger reuse factor of pilot sequences.
After observing the received pilot signal,
which is transmitted from the scheduled UE and is interfered by the other UEs in neighbouring cells using the same pilot sequence,
a BS can detect the corresponding pilot and then estimate the channel by, e.g., a minimum mean square error (MMSE) estimator.

\subsection{Wireless System Model}

The 3D distance between an arbitrary UE and an arbitrary BS is denoted as
\begin{equation}
	w=\sqrt{r^{2}+h^{2}},
\end{equation}
where $r$ is the 2D distance between an arbitrary UE and an arbitrary BS,
and $h$ is the absolute antenna height difference between these two.
Note that the value of $h$ is in the order of several meters.
In our simulation, the height difference is decided according to the path loss model~\cite{Ding2016GC_ASECrash,3GPP}.

With regard to such path loss modelling,
we adopt a general and practical piecewise path loss model with respect to the 3D distance $w$ proposed in~\cite{PL model},
where each segment of the path loss function is modelled as either a LoS transmission or a NLoS one.
Such path loss function is given by
\begin{equation}
\zeta^{\textrm{}}_{jlk}(w)=
    \begin{cases}
    \zeta_{1(jlk)}^{\textrm{}}(w), & ~ \text{when}~ h \leq w \leq d_1^{\textrm{}}\\
    \zeta_{2(jlk)}^{\textrm{}}(w),  &~  \text{when}~d_1^{\textrm{}} \leq w \leq d_2^{\textrm{}}\\
\vdots & \vdots\\
    \zeta_{n(jlk)}^{\textrm{}}(w),  &~  \text{when}~w\geq d_N^{\textrm{}}\\
    \end{cases},
\end{equation}
where $\zeta_{n(jlk)}^{\textrm{}}(w)$ is the $\emph{n}$-th segment of the path loss function between the $\emph{j}$-th BS and the $\emph{k}$-th UE scheduled by the $\emph{l}$-th BS.

Each segment of path loss function is modelled by:
 \begin{equation}
 \zeta_{n}^{\textrm{}}(w)=
    \begin{cases}
    \zeta_{n}^{\textrm{L}}=A_n^{\textrm{L}} w^{-\alpha_n^{\textrm{L}}}, &  \text{for LoS}\\
    \zeta_{n}^{\textrm{NL}}=A_n^{\textrm{NL}} w^{-\alpha_n^{\textrm{NL}}}, &  \text{for NLoS}\\
    \end{cases},
 \end{equation}
where
\begin{itemize}
\item
$\zeta_{n}^{{\rm {L}}}\left(w\right)$ and $\zeta_{n}^{{\rm {NL}}}\left(w\right),n\in\left\{ 1,2,\ldots,N\right\} $ are the $n$-th segment of the path loss functions for the LoS and the NLoS cases, respectively,
\item
$A_{n}^{{\rm {L}}}$ and $A_{n}^{{\rm {NL}}}$ are the path loss values at a reference 3D distance $w=1$ for the LoS and the NLoS cases, respectively, and
\item
$\alpha_{n}^{{\rm {L}}}$ and $\alpha_{n}^{{\rm {NL}}}$ are the path loss exponents for the LoS and the NLoS cases, respectively.
\end{itemize}


In addition, the piecewise probability function of that a transmitter and a receiver communicate via a LoS path while separated by a distance $w$ is written as
\begin{equation}
Pr^{\textrm{L}}=
    \begin{cases}
    Pr_1^{\textrm{L}}(w)  & ~ \text{when}~ h \leq w \leq d_1^{}\\
    Pr_2^{\textrm{L}}(w)  & ~ \text{when}~ d_1^{} \leq w \leq d_2^{}\\
\vdots & \vdots\\
    Pr_n^{\textrm{L}}(w)  & ~ \text{when}~ w \geq d_N^{}\\
    \end{cases},
\end{equation}
where $Pr_n^{\textrm{L}}(w)$ is the $\emph{n}$-th segment of the LoS probability function that the path between an arbitrary UE and an arbitrary BS is a LoS one.

Based on these path loss and probabilistic LoS/NLoS models,
a practical UE association strategy (UAS) is considered in this paper.
A UE is associated with the BS that provides the maximum average received signal strength
(i.e. the largest $\zeta(w)$).
Moreover, we assume that one $M$-antenna BS can at most simultaneously schedule $K_{\textrm{U}}$ UEs in a time-frequency resource block according to~\cite{MassiveMK},
where $K_\textrm{U}$ is given by
\begin{equation}
	K_{\textrm{U}}=min\left\{K_\textrm{T},M/4\right\},
\end{equation}
where $K_\textrm{T}$ is the number of pilot sequences and $M$ is the number of antennas per BS.
Note that $M/4$ is an empirical value to achieve a good performance for a MU-MIMO system in case of pilot contamination~\cite{MassiveMK}.

For convenience,
we denote a UE in each BS by the index of its used uplink training sequence,
i.e., the $k$-th UE is the UE that uses the $k$-th uplink training sequence $t_k$,
where $k$ is randomly chosen from 1 to the maximum training sequence number $K_{T}$.
Note that the $k$-th UE could be an empty UE since not all of the uplink training sequences are used up in each BS.
If there are more than $K_{\textrm{U}}$ UEs connected to a BS,
only up to $K_{\textrm{U}}$ of them are randomly chosen to be scheduled,
which means that this BS is fully loaded and $K_{\textrm{U}}$ training sequences are used.
Without loss of generality  and as mentioned earlier,
we consider the first UE in $B_1$ as the typical UE,
denoted by $U_{1}$.

From~\cite{UEdistribution},
the number of UEs per active BS can be modelled as a Negative Binomial distribution,
i.e., $K\sim$$\cal{NB}$$(q,\frac{\rho}{\rho+q\lambda})$.
However, note that only active BSs are considered to participate in DL transmissions and that at most $K_{\textrm{U}}$ UEs can be served simultaneously by an active BS.
Thus, the actual number of scheduled UE per active BS $\hat{K}$ can be modelled as a truncated modified Negative Binomial distribution,
i.e., $\hat{K}\sim$$Trunc$$\cal{NB}^{*}$$(q,\frac{\rho}{\rho+q\lambda})$,
and the PMF of $\hat{K}$ can be expressed as
\begin{equation}
    f_{\hat{K}}(k)=
        \begin{cases}
            \frac{f_K(k)}{1-f_K(0)} \qquad 1\leq k \leq K_{U}-1\\
            \frac{\sum_{k=K_{U}}^{\infty}f_K(k)}{1-f_K(0)} \qquad k\doteq K_{U}\\
        \end{cases},
\end{equation}
where $f_K(k)$ is the PMF of UE number distribution (i.e., $K\sim$$\cal{NB}$$(q,\frac{\rho}{\rho+q\lambda})$) derived in~\cite{UEdistribution},
which is given by
\begin{eqnarray}
f_{K}\left(k\right)\hspace{-0.2cm} & = & \hspace{-0.2cm}\Pr\left[K=k\right]\nonumber \\
\hspace{-0.2cm} & \overset{}{=} & \hspace{-0.2cm}\frac{\Gamma(k+q)}{\Gamma(k+1)\Gamma(q)}\left(\frac{\rho}{\rho+q\lambda}\right)^{k}\left(\frac{q\lambda}{\rho+q\lambda}\right)^{q}.\label{eq:NB_PMF}
\end{eqnarray}
where $\text{\ensuremath{\Gamma}}(\cdot)$ is the Gamma function.
Note that $f_{K}\left(k\right)$ satisfies the normalization condition:
$\sum_{k=0}^{+\infty}f_{K}\left(k\right)=1$.
\vspace{0cm}

\subsection{Pilot-Aided mMIMO Channel Estimation}

The channel is assumed to be invariant in a time-frequency resource block,
and change independently from block to block.
The channel vector can be expressed as
\begin{equation}
\textbf{h}_{jlk}^{{(PL)}}=(\zeta_{jlk}^{{(PL)}})^{\frac{1}{2}}{\phi}_{jlk}^{\frac{1}{2}}\textbf{w}_{jlk},
\end{equation}
where $\emph{'\textrm{PL}'}$ takes the value 'L' and 'NL' for LoS transmissions and NLoS transmissions, respectively,
$\textbf{h}_{jlk}^{{(PL)}}$ is the channel vector between the $\emph{j}$-th BS and the $\emph{k}$-th UE scheduled by the $\emph{l}$-th BS,
$\textbf{w}_{jlk}$ is the multi-path fading vector modelled according to Rayleigh fading, and
$\phi_{jlk}$ is the covariance matrix of the channel.
Since we consider that the channel $\textbf{h}_{jlk}^{(PL)}$ is identically and independently distributed (i.i.d.),
$\phi_{jlk}$ should be an identity matrix in this case.

In the uplink training stage,
pilot contamination is considered in our simulation.
Based on the distance-dependent fractional power compensation scheme,
the channel vector $\textbf{y}_{11}^{{(PL)}}$ observed at BS $B_1$ for the typical UE $U_1$ is given by
\begin{equation}
	\textbf{y}_{11}^{{(PL)}}=\sqrt{P_{11}^{\textrm{tx}}}\textbf{h}_{111}^{{(PL)}}+
	\sum_{l\neq1}\sqrt{P_{l1}^{\textrm{tx}}}\textbf{h}_{1l1}^{{(PL)}}+\bf{n}_\textrm{11},
 \end{equation}
where $l$ represents the $l$-th BS,
which serves an interfering UE using the first pilot sequence.
Besides, $\bf{n}_\textrm{11}$ denotes a zero-mean additive white Gaussian noise (AWGN) vector at the typical UE $U_1$,
where the variance of each element is $\sigma^2$.

Using the fractional power compensation~\cite{3GPP},
$P_{lk}^{\textrm{tx}}$ is the transmit power from the $\emph{k}$-th UE in the $\emph{l}$-th BS,
which can be expressed as
\begin{equation}
	P_{lk}^{\textrm{tx}}=P^{\textrm{tx}}_{\textrm{u}}(\zeta^{(PL)}_{llk}(w))^{-\epsilon },
\end{equation}
where $P^{\textrm{tx}}_{\textrm{u}}$ is the baseline transmit power of each UE,
and $\epsilon$ is the fraction of the path loss compensation.

From the observation of the pilot signals transmitted from UEs,
BSs can estimate their channels by correlating the corresponding pilot sequences with the observation by using an MMSE estimator.
Since the channel $\textbf{h}_{111}^{{(PL)}}$ is modelled as i.i.d Rayleigh fading,
the estimated channel $\bar{ \textbf{h}}_{111}^{{(PL)}}$ can be calculated as
\begin{equation}
	\bar{ \textbf{h}}_{111}^{{(PL)}}=\frac{\sqrt{P_{1  1}^{\textrm{tx}}}\zeta_{111}^{{(PL)}}}
	{\sum_{l\neq1}P_{l1}^{\textrm{tx}}\zeta^{{(PL)}}_{1l1}+\sigma^2 }\textbf{y}_{11}^{{(PL)}}.
\end{equation}

Note that the estimation error of $\bar{ \textbf{h}}_{111}^{(\textrm{PL)}}$ can be formulated by
\begin{equation}
	\hat{ \textbf{h}}_{111}^{{(PL)}}=\textbf{h}_{111}^{{(PL)}}-\bar{ \textbf{h}}_{111}^{{(PL)}}.
\end{equation}

\vspace{0cm}

\subsection{Performance Metrics}

In the downlink data transmission stage,
we assume that BSs apply the zero-forcing (ZF) precoder based on the estimated channel obtained in the uplink training stage.
We assume that the total power of a BS is fixed,
and it is equally divided among the served UEs.

With these assumptions,
the received symbol $\textbf{s}_{1}$ at typical UE $U_{1}$ can be written as~\cite{Ding2013MUMIMO}
\begin{equation}
	\begin{split}
	\textbf{s}_{1}= &\sqrt{P_{\textrm{1}}}\bar{\textbf{h}}_{111}^{{(PL)}H}\textbf{f}_{11}^{{}}s_{11}+
	\sqrt{P_{\textrm{1}}}\hat{\textbf{h}}_{111}^{{(PL)}H}\textbf{f}_{11}^{{}}s_{11}\\&
	+\sqrt{P_{\textrm{l}}}\sum_{(l,k)\neq (1,1)} \textbf{h}_{1lk}^{{(PL)}H}\textbf{f}_{lk}^{}s_{lk}+
	\textbf{n}_1,
	\end{split}
\end{equation}
where $\textbf{f}_{lk}^{{}}$ is the ZF precoding vector for the $k$-th UE scheduled by the $l$-th BS,
$P_{\textrm{l}}$ is the transmit power allocated by the $l$-th BS to each of its scheduled UE,
$s_{lk}$ is the signal intended for $k$-th UE in the $l$-th BS, and
$\textbf{n}_1$ is the AWGN vector at the typical UE $U_1$. Here, $\textbf{f}_{lk}^{{}}$ is computed by:
\begin{equation}
    \mathrm{\mathbf{f}}_{lk}^{}=\frac{\left(\bar{\mathbf{h}}_{llk}^{\mathrm{\mathit{\left(PL\right)}}H}\bar{\mathbf{h}}_{llk}^{\mathrm{\mathit{\left(PL\right)}}}\right)^{-1}\bar{\mathbf{h}}_{llk}^{\mathrm{\mathit{\left(PL\right)}}H}}{\mathcal{\mathbb{E}}\left\{\left|\left(\bar{\mathbf{h}}_{llk}^{\mathrm{\mathit{\left(PL\right)}}H}\bar{\mathbf{h}}_{llk}^{\mathrm{\mathit{\left(PL\right)}}}\right)^{-1}\bar{\mathbf{h}}_{llk}^{\mathrm{\mathit{\left(PL\right)}}H}\right|^{2}\right\}}.
\end{equation}

In (14),
the first term is the received signal and the others are unknown at the UE.
Hence,
the downlink SINR at the typical UE $U_1$ can be calculated by
\begin{equation}
	\begin{split}
	&\textrm{SINR}=\\
	&\frac{P_1{\left|\bar{ \textbf{h}}_{111}^{{(PL)}H}\textbf{f}_{11}^{{}}\right|}^2}{P_{1}{\left|\hat {\textbf{h}}_{111}^{{(PL)}H}\textbf{f}_{11}^{{}}\right|}^2+\sum_{(l,k)\neq (1,1)} P_{1}{\left|\textbf{h}_{1lk}	^{{(PL)H}}\textbf{f}_{lk}^{{}}\right|}^2+\sigma^2 }.
	\end{split}
\end{equation}

We also investigate the area spectral efficiency (ASE) performance
in $\textrm{bps/Hz/km}^{2}$, which is defined as
\begin{equation}
A^{{\rm {ASE}}}\left(\gamma_{0}\right)=\tilde{\lambda}\int_{\gamma_{0}}^{+\infty}\log_{2}\left(1+\gamma\right)f_{\mathit{\Gamma}}\left(\gamma\right)d\gamma,\label{eq:ASE_def}
\end{equation}
where $\gamma_{0}$ is the minimum working SINR in a practical SCN,
and $f_{\mathit{\Gamma}}\left(\lambda,\gamma\right)$ is the PDF of
the SINR observed at the typical UE for a particular value
of $\lambda$.
\vspace{0cm}

\section{Results and discussion}

In this section,
we investigate the ASE performance with various network deployment strategies via simulations.
We consider a practical two-piece path loss function and a practical two-piece exponential LoS probability function,
defined by~\cite{3GPP}.
In more detail,
all simulation parameters are taken from Tables A.1-3, A.1-4, A.1-5 and A.1-7 of \cite{3GPP}:
$\alpha^{{\rm {L}}}=2.09$, $\alpha^{{\rm {NL}}}=3.75$, $A^{{\rm {L}}}=10^{-10.38}$, $A^{{\rm {NL}}}=10^{-14.54}$, $P^{\textrm{tx}}=24$\ dBm, $P_{{\rm {N}}}=-95$\ dBm (including a noise figure of 9\ dB at each UE), $e=1$.
In our simulations, the total number of antennas per square kilometre is set to 500\,antennas/km$^2$ and 1000\,antennas/km$^2$ respectively,
which are practical assumptions for 5G~\cite{Tutor_smallcell}.
In addition,
we set the total number of uplink training sequences to $K_\textrm{T}=20$, which are reused across all cells.
Note that the pilot reuse factor varies with the BS density and UE density in different network deployment. More specifically, the pilot reuse factor is higher in a network with a larger UE density or a smaller BS density since the UE number scheduled by each BS becomes larger. As a result, the UE density generating the pilot contamination is the product of the pilot reuse factor and the BS density.


\vspace{0cm}

\subsection{The ASE Performance}

Figs.~1 and 2 show the ASE performance versus the UE density $\rho$ for various BS densities,
when the antenna densities are 500 antennas/km$^2$ and 1000 antennas/km$^2$, respectively.
Note that only UEs whose SINR is larger than $\gamma_0=0$\,dB are considered in the computation of the ASE according to (17).

\begin{figure}
	\centering
	\includegraphics[width = 8cm]{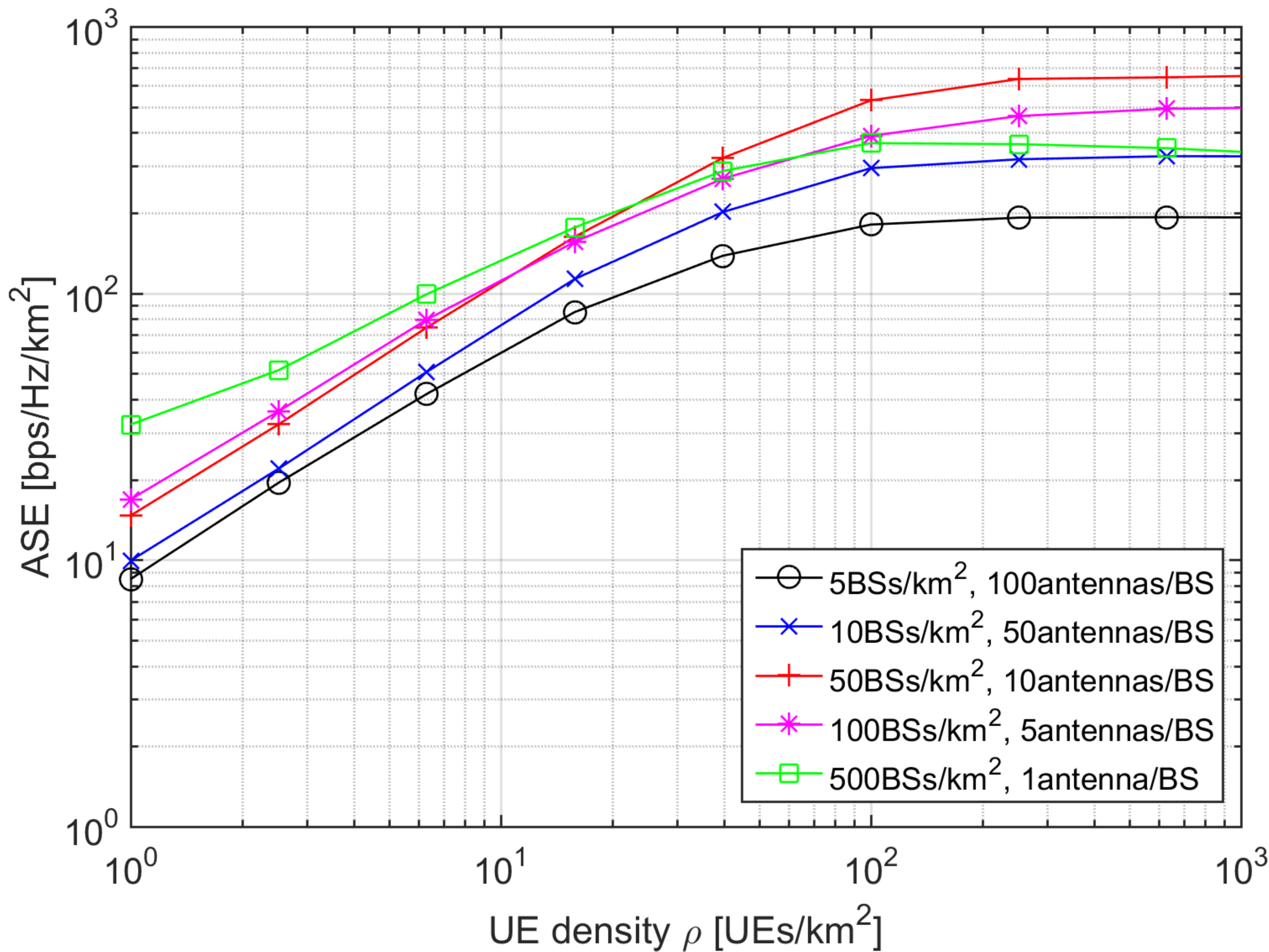}
	\caption{The ASE performance of different antenna deployment when $\gamma_0=0$\,dB vs. UE density $\rho$ with 500\,antennas/km$^2$} \label{figure1}
\end{figure}

\begin{figure}
	\centering
	\includegraphics[width = 8cm]{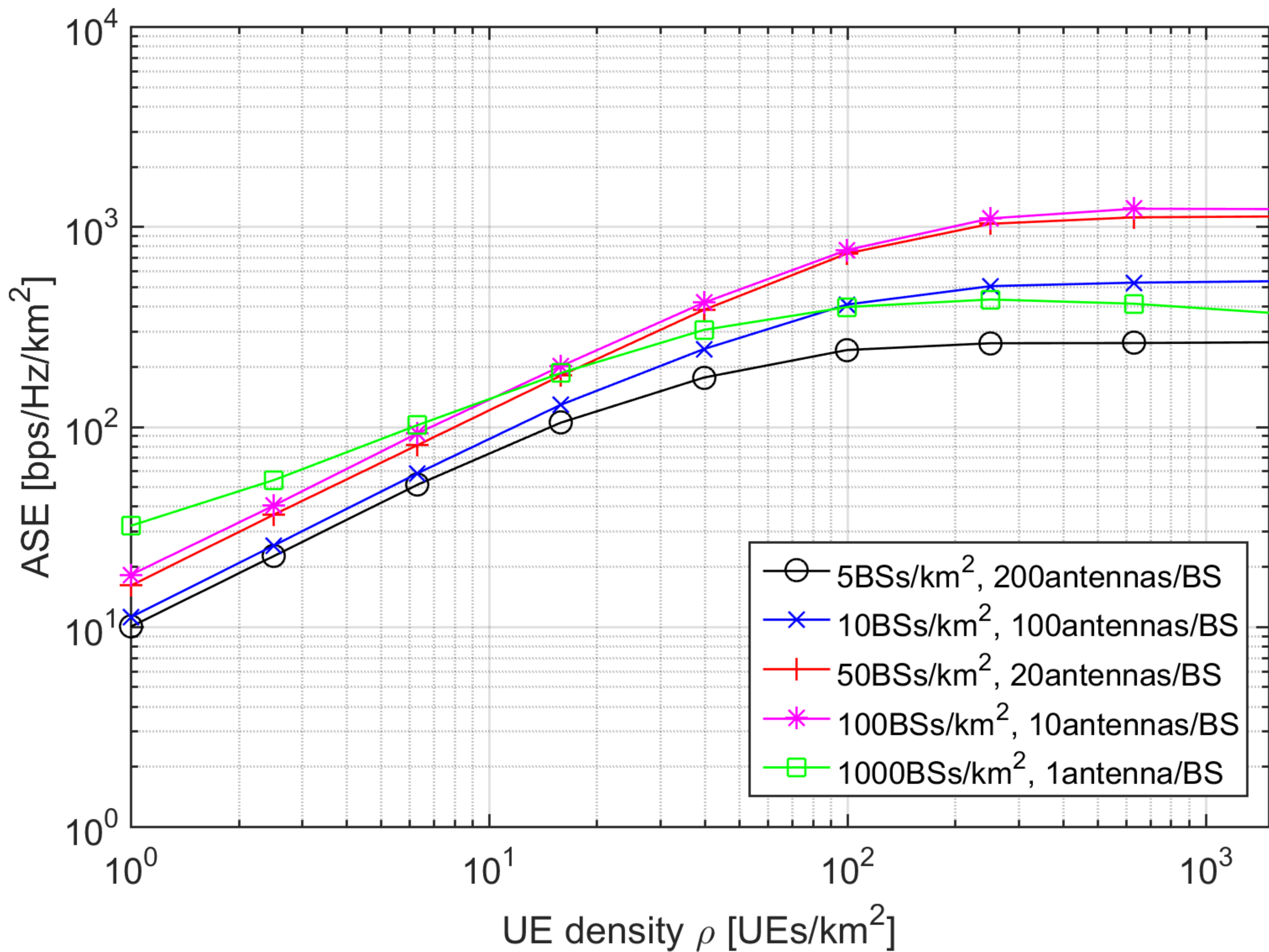}
	\caption{The ASE performance of different antenna deployment when $\gamma_0=0$\,dB vs. UE density $\rho$ with 1000\,antennas/km$^2$} \label{figure1}
\vspace{-0.2cm}
\end{figure}

From Figs.~1 and 2,
we can draw the following conclusions:
\begin{itemize}

\item
Network deployments with BS density $\lambda$ at 50 and 100 BSs/km$^2$ always perform better than those at 5 and 10 BS/km$^2$ in term of ASE.
For this BS density range,
this means that a dense network with many BSs and a few antennas per BS is a better solution than a sparse network with a few BSs and many antennas per BS.
The intuition behind this phenomenon is that by having more BSs,
the power gain due to the short signal-link distance surpasses the loss of beamforming gain caused by the less antennas per BS.

\item
Note that the single antenna cases,
i.e., 500\,BSs/km$^2$ in Fig. 1 and 1000\,BSs/km$^2$ in Fig. 2,
achieves their highest ASE performance when $\rho$ is around 100\,UEs/km$^2$ and 200\,UEs/km$^2$, respectively.
This is because more BSs are activated in a scenario with more UEs,
which in turn causes a large number of interference paths to transition from NLoS to LoS,
which damages the overall network performance~\cite{PLmodel}.
Moreover, as shown in Figs. 1 and 2,
the single antenna case show a better performance than the other investigated cases at around $\rho \in [1,20]$\,UEs/km$^2$ and $[1,10]$\,UEs/km$^2$, respectively.
The reason is that when $\rho$ is relatively small,
the active BS density $\tilde{\lambda}$ in each simulated case is almost the same,
i.e., $\rho$,
and thus bringing the limited number of active BSs closer to the UEs is a better strategy than the mMIMO one.

\item
For the multiple-antenna deployment strategies with the BS density much less than 500\,BSs/km$^2$ in Fig. 1 and 1000\,BSs/km$^2$ in Fig. 2,
the ASE firstly grows dramatically as the UE density $\rho$ increases,
and then suffers from a slow growth when the UE density $\rho$ is larger than a threshold.
The reason is that BSs are low loaded or even deactivated when the UE density $\rho$ is small.
Thus, every newly added UE can be scheduled with a decent link quality,
making a significant contribution to the ASE.
In contrast,
nearly all BSs are active and serving the maximum number of UEs when $\rho$ is large enough.
Hence,
it is not very likely that a newly added UE can get scheduled with an additional degree of freedom.
The worst case scenario is when BSs are fully loaded,
i.e., the UE density $\rho$ is large enough compared to the BS density $\lambda$,
which results in a saturated ASE as shown in Figs. 1 and 2.

\item
From the ASE results shown in Fig.1 and Fig. 2,
there exists an optimal network deployment strategy for a specific antenna density and UE density $\rho$.
In more detail,
when the antenna density is 500 antennas/km$^2$ and $\rho$ is 600\,UEs/km$^2$,
the descending order of the ASE performance is 50, 100, 500, 10 and 5\,BSs/km$^2$.
This means that the optimal BS density lies between 10\,BSs/km$^2$ and 100\,BSs/km$^2$ for the case of 500 antennas/km$^2$ and $\rho=600\,\textrm{UEs/km}^2$.
Moreover,
when the antenna density is 1000 antennas/km$^2$ and $\rho$ is 600\,UEs/km$^2$,
the descending order of the ASE performance is 100, 50, 10, 500 and 5\,BSs/km$^2$.
This means that the optimal BS density lies between 50\,BSs/km$^2$ and 500\,BSs/km$^2$ for the 1000 antennas/km$^2$ case when $\rho=600\,\textrm{UEs/km}^2$.
\end{itemize}

\vspace{0cm}

\subsection{The Trade-off between the BS Density and the Antenna number per BS }

Fig. 3 shows the ASE performance versus the BS density $\lambda$ for various UE densities.
In particular,
we consider four UE densities:
$\rho=50$\,UEs/km$^2$, $\rho=100$\,UEs/km$^2$, $\rho=300$\,UEs/km$^2$ and $\rho=600$\,UEs/km$^2$ and a total antenna density of 1000\,antennas/km$^2$.
To evaluate the impact of different UE densities on the ASE performance,
we keep the other assumptions and models same as before.

\begin{figure}
	\centering
	\includegraphics[width = 8cm]{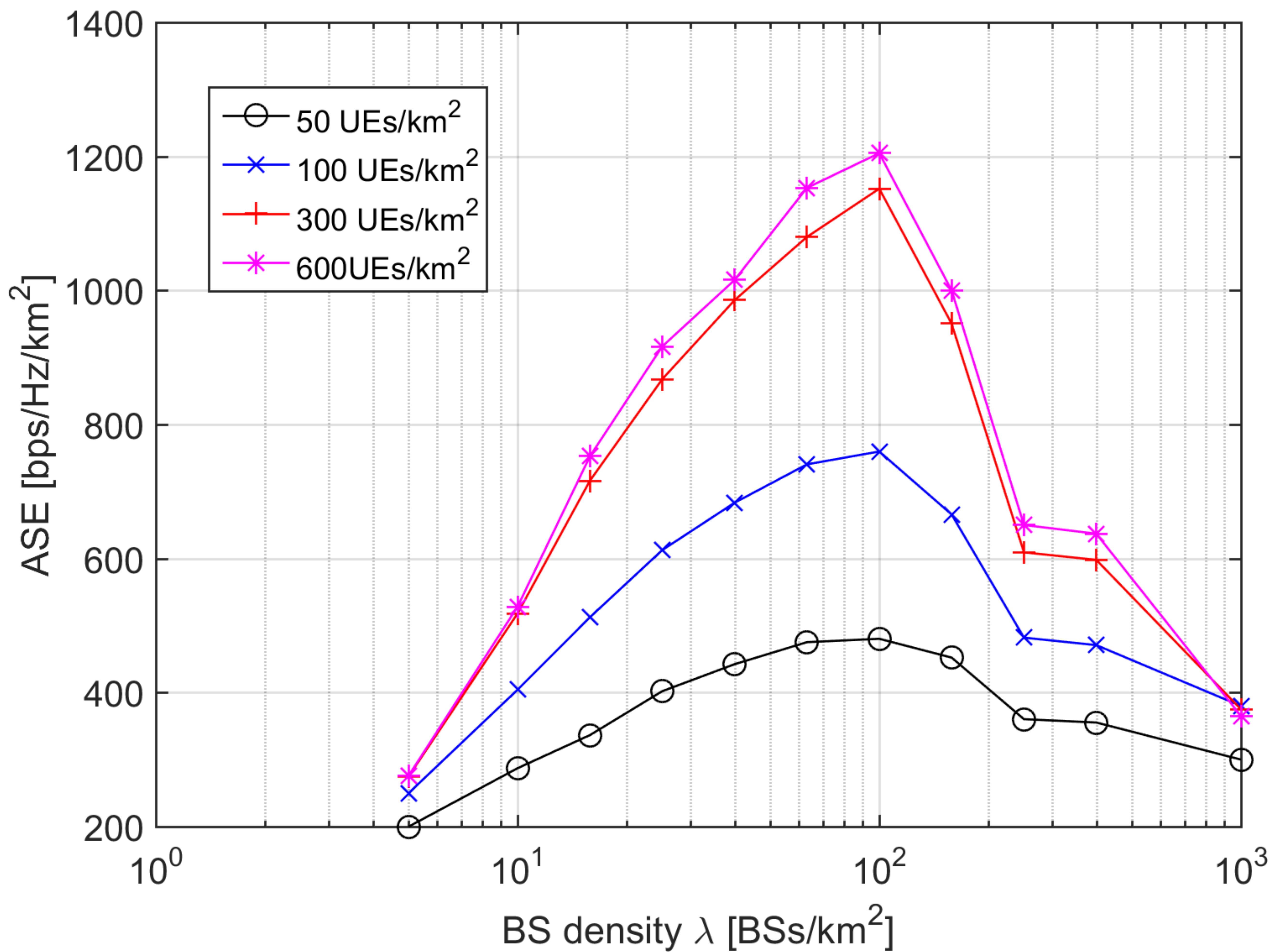}
	\caption{The ASE performance of different UE density when $\gamma_0=0$\,dB vs. BS density $\lambda$ with 1000\,antennas/km$^2$} \label{figure2}
\vspace{-0.2cm}
\end{figure}

From Fig. 3, we can observe that:
\begin{itemize}

\item
For a fixed antenna density (1000\,antennas/km$^2$) and all the investigated UE densities,
there exists an optimal network deployment strategy to maximize the ASE.
The optimal deployment is around the BS density $\lambda=100$\,BSs/km$^2$ with approximately 10 antennas per BS
for these UE density.
The intuition is that
\emph{(i)} by densifying the network with more BSs,
we can achieve a receive power gain due to the smaller distance between the typical UE and its serving BS;
\emph{(ii)} by installing more antennas on each BS,
we can achieve a beamforming gain for UEs using mMIMO,
although such beamforming gain is degraded by pilot contamination.
Thus, a trade-off exists between the receive power gain and the beamforming gain,
if we fix the antenna density in the network.

\item
It is important to note that the larger the UE density, the better the ASE.
This is because a larger UE density results in more UEs scheduled per square kilometer.
Also note that the ASE performance difference between the $\rho=600$\,UEs/km$^2$ case and the $\rho=300$\,UEs/km$^2$ case is smaller than that between the $\rho=300$\,UEs/km$^2$ case and the $\rho=100$\,UEs/km$^2$ case.
This is in line with the ASE trend shown in Fig. 1,
which increases rapidly and then suffers from a slow growth with the UE density due to the performance saturation.

\item
For the investigated antenna density (1000\,antennas/km$^2$),
the optimal BS densities for various UE densities are the same,
i.e., around $\lambda=100$\,BSs/km$^2$,
which indicates that the optimal network deployment might be independent of the UE density.
This conjecture needs to be further studied with theoretical analysis.
\item
Note that the ASE experiences a slow decrease when BS density is larger than around 250 BSs/km$^2$. The reason is that the maximum scheduled UE number per BS has already decreased to one according to (6), which means the ASE performance can not be further influenced by decreasing the scheduled UE number per BS.
\end{itemize}

\vspace{0cm}

\subsection{Performance Impact of Pilot Contamination}

An interesting question follows from Fig. 3 is whether the trade-off still exists between the receive power gain and the beamforming gain,
if the pilot contamination is removed from the investigated MU-MIMO network (i.e., perfect channel state information).
To answer this question, in Fig. 4 we plot the ASE performance without pilot contamination,
while keeping the other assumptions same as those for Fig. 3.
\begin{figure}
	\centering
	\includegraphics[width = 8cm]{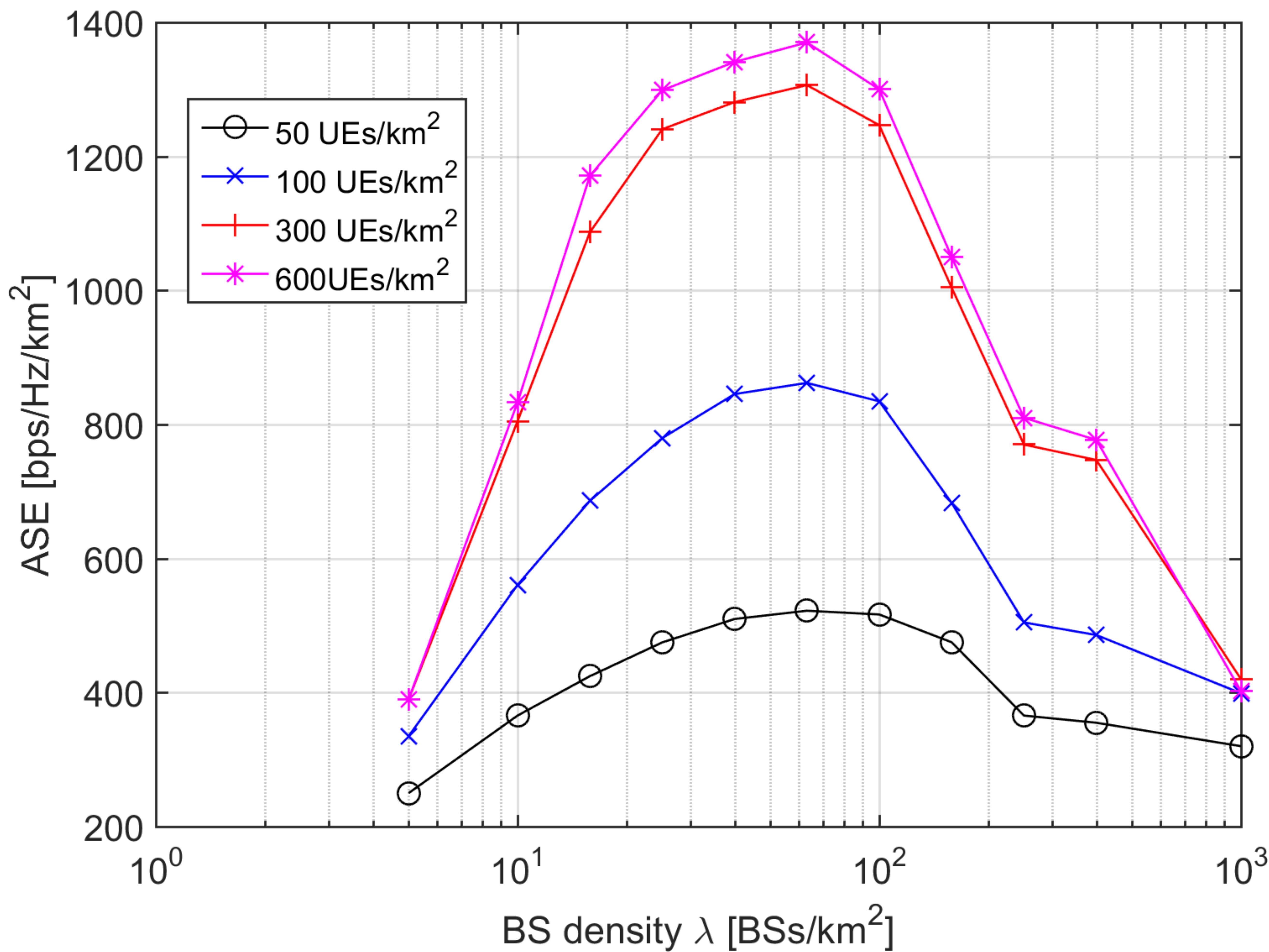}
	\caption{The ASE performance without pilot contamination of different UE density when $\gamma_0=0$\,dB vs. BS density $\lambda$ with 1000\,antennas/km$^2$} \label{figure2}
\vspace{-0.5cm}
\end{figure}

From Fig. 4, we can conclude that:
\begin{itemize}
\item
Without pilot contamination,
the ASE performance improves for all cases,
thanks to the accurate CSI.
\item
Meanwhile,
the descending order of the ASE performance does not change if we remove the pilot contamination,
which means that the trade-off still exists between the receive power gain and the beamforming gain,
regardless of the pilot contamination phenomenon.
Also note that without pilot contamination, the optimal network deployment shifts to the BS density around 63 BSs/km$^2$ with 16 antennas per BS.
\item
The ASE performance improvement without pilot contamination is small when the BS density is relatively large,
because the pilot reuse factor is small in dense networks due to the limited number of UEs per BS.

\end{itemize}

\vspace{0cm}

\section{Conclusion}
In this paper,
we have conducted a performance evaluation with a fixed number of antennas per square kilometer.
Our results indicate that there exists an optimal network deployment strategy to maximise the ASE performance for a certain UE density.
Intuitively speaking, a balance between
\emph{(i)} bringing UEs closer to BSs by densifying the network, and
\emph{(ii)} allowing for more antennas per BS to achieve a higher precoding gain with the consideration of pilot contamination,
needs to be found to optimise the system performance.

\vspace{0.3cm}

\bibliographystyle {IEEEtran} 
\bibliography{ref} 
\end{document}